\documentclass[aps,prd,superscriptaddress,nofootinbib,eqsecnum,twocolumn]{revtex4-1}

\pdfoutput=1

\usepackage{graphicx}  
\usepackage{dcolumn}   
\usepackage{amssymb}   
\usepackage{xcolor}
\usepackage{graphics} 
\usepackage{hyperref}
\usepackage{amsfonts}
\usepackage{latexsym}
\usepackage{slashed}
\usepackage{amsmath}
\usepackage{amsfonts}
\usepackage{amssymb}
\usepackage{dsfont}
\usepackage{bm}  
\usepackage[normalem]{ulem} 
\usepackage{amsmath,mathtools,mathrsfs,bbm,amssymb} 
\usepackage{bbold}
\usepackage{subfigure}
\usepackage{array}
\usepackage{wrapfig}
\usepackage{multirow}
\usepackage{tabularx}
\usepackage{float}
\usepackage{array}
\usepackage{booktabs}
\usepackage{commath}
\usepackage{graphicx}
\usepackage{epstopdf}
\usepackage{amsmath}
\usepackage{amsfonts}
\usepackage{amssymb}
\usepackage{dsfont}
\usepackage{latexsym}
\usepackage{hyperref}
\usepackage[english]{babel}
\usepackage[utf8]{inputenc}
\usepackage[colorinlistoftodos]{todonotes}
\usepackage{xcolor}
\usepackage{slashed}
\usepackage{bm}  
\usepackage[normalem]{ulem} 
\usepackage{amsmath,mathtools}
\usepackage{caption}
\usepackage{subcaption}
\usepackage{float}
\usepackage{MnSymbol}
\useunder{\uline}{\ul}{}
\usepackage{hyperref}
\begin{document}

\title{Spontaneous collapse effects on relativistic fermionic matter}

\author{Y. M. P. Gomes}
\email{ymuller@cbpf.br}
 \affiliation{Centro Brasileiro de Pesquisas F\'isicas, Rua Dr. Xavier Sigaud 150, Urca, CEP: 22290-180, Rio de Janeiro-RJ, Brazil}


\begin{abstract}
This study expands the spontaneous collapse assumptions into the relativistic quantum field theory framework for Dirac fields. By solving Lindblad's master equation using the Keldysh formalism, the effective action is derived, which captures the dynamics of fermions with spontaneous collapse represented as an imaginary self-interaction term. Utilizing the corresponding Dyson-Schwinger equations at 1-loop approximation, the effective mass induced by the nonlinearity is computed. The findings indicate the presence of a new mechanism that introduces a qualitative change in the mass spectrum, where the particle's mass becomes complex. Based on the stability of the proton, new bounds on the parameters are found. This mechanism also generates a Lorentz invariance violation in the infrared regime and recovers the Lorentz invariance in the ultraviolet regime. The corresponding hydrodynamics of the system is analyzed through the Keldysh component of the propagator, and a conserved charge is found. In contrast, the energy-momentum tensor is shown to be non-conserving. This phenomenon represents a new contribution to the understanding of the spontaneous collapse and the transition from quantum to the classical realm. 
\end{abstract}


\maketitle

\section{Introduction}
Classical mechanics and electromagnetism reached their limits at the end of the 19th century, when quantum mechanics, formulated by Bohr, Heisenberg, Schrödinger, and many others, was shown to be the most refined overcoming of the previous physical and philosophical standpoint, providing an extremely successful explanation of atomic phenomena. However, it also gave rise to several paradoxes that are still debated today. One of such paradoxes is represented by Bohr's ``Principle of Complementarity" \cite{bohr1}. This principle states that there are complementary variables in microphysics that cannot be measured (or known) simultaneously, serving as the general basis for Heisenberg’s uncertainty principle. This principle underpins the so-called Copenhagen interpretation, a central point of criticism against quantum mechanics, particularly from followers of the school founded by David Bohm and de Broglie \cite{bohm1}. The Bohmian theory is deterministic, with randomness entering, like in classical mechanics, due to the initial conditions. Another important feature of the Copenhagen and pilot-wave interpretations is the superposition of macroscopic states\cite{bohm2}. In addition to the controversy surrounding the ``Principle of Complementarity", there is also the debate over the wave function collapse, illustrated by Schrödinger's cat paradox. Quantum mechanics has successfully explained microscopic phenomena, as evidenced by over a century of experimental validation and the many technologies it has enabled. Nevertheless, there is still no consistent theory that explains the dynamics of wave function collapse.

The Copenhagen interpretation postulates an artificial division between the micro- and macro-worlds without quantitatively specifying at what mass scale this division should occur. In this perspective, microscopic objects obey the rules of quantum theory, while macroscopic objects obey classical mechanics. During the measurement process, when a microsystem interacts with a macrosystem, the wave function of the microsystem ``collapses" to one of the eigenstates of the measured observable. It is postulated that this collapse occurs by the Born probability rule, and no dynamic mechanism is specified to explain how the collapse occurs \cite{sakurai}. In particular, it is known that this collapse process contradicts the linearity of the Schrödinger equation, as the collapse is a process that does not preserve the purity of the previous state, an intrinsic property of Schrödinger's dynamics. Succinctly, we can say that, according to this interpretation, evolution in quantum theory occurs in two ways: deterministic evolution, according to Schrödinger's equation, before and after a measurement, and non-deterministic, probabilistic evolution during a measurement.

Based on this inconsistency, several authors propose that the Schrödinger equation should be modified so that superpositions of macroscopically different states, as didactically exemplified by Schrödinger's cat, cannot arise or persist enough to be detected. Theories like these are commonly referred to as spontaneous collapse (SC) models. Currently several non-relativistic models, as the Ghirardi-Rimini-Weber theory (GRW) \cite{GRWref} being the first attempt, followed by the Diosi-Penrose (DP) models \cite{DPref1,DPref2}, and the Continuous Spontaneous Localization (CSL) \cite{CSLref}, have proven capable of start to address the issue of macroscopic superposition, as the non-linear dynamics of these models generate a spontaneous measurement of the mass density operator $m({\bf x},t)$ at all ${\bf x}$ and $t$. The SC models have also cosmological implications in the quantum-to-classical transition of the Universe \cite{Reyes24}.

The present work focuses on the CSL model as a representative model of spontaneous collapse. One important feature of the CSL-like models is given by the amplification mechanism, which is provided by the fact that the collapse rate of the object is proportional to its size or mass. 
Consequently, contrary to Copenhagen and Bohmian perspectives, the persistence of (macroscopic) superpositions is lost at the cost of modifying unitary dynamics. Considering the observables of the models, the corresponding dynamical equation for the statistical operator $\hat{\rho}$ is a Lindblad master equation commonly used to describe Markovian open quantum systems. In this context, the system's dynamical evolution ceases to be unitary while still maintaining trace preservation and ensuring the positivity of the density matrix \cite{lindref1}, \cite{lindref2}.

The Lindblad formalism is primarily used to describe non-relativistic systems \cite{lindexample1}-\cite{lindexample4}, particularly in condensed matter physics. Recently, studies have explored its potential application to relativistic systems, specifically in the bosonic \cite{lindref3},\cite{lindref4}, and fermionic case \cite{gomes24}. These investigations reveal that the presence of dissipation generates an explicit violation of the manifest Lorentz invariance.  Thus, the Lorentz violation characteristic of the SC models can be seen as a way to impose bounds on the parameters and a possible window to better understand the manifestation of the collapse of the wave function on relativistic systems.

This work is organized as follows: Section 2 introduces the main aspects of spontaneous-collapse models within the Keldysh formalism and presents the action of the system. In Section III, by use of the path-integral technique, the Dyson-Schwinger (DS) equations are reached, and a generalization of the collisionless Boltzmann equation is found. With a truncation of the Dyson series, the solutions of the DS equations are found. In Section 4, the results are shown, and in Section 5, the final comments and perspectives are presented. This paper will consider the natural units where $\hbar = k_b = c = 1$. In the paper, $[A,B] = AB - BA$ and $\{A,B\} = AB + BA$, and the spacetime have $D =3+1$ dimensions.

\section{Continuous spontaneous localization (CSL) model }
The Schrödinger equation for a relativistic many-body (fermionic) system can be written as $ d| \psi_t \rangle = - i\hat{H}  | \psi_t \rangle dt $, where the Hamiltonian operator is given by $\hat{H} = \sum_{s, {\bf p}} \epsilon_{\bf p}\left( a_{s,{\bf p} }^\dagger a_{s,{\bf p}}+b_{s,{\bf p}}^\dagger b_{s,{\bf p}}\right) $, with $a^\dagger_{s,{\bf p}}, a_{s,{\bf p}}$ and $b^\dagger_{s,{\bf p}},b_{s,{\bf p}}$ the creation/anihilation operator for the particle and anti-particles, respectively, and $\epsilon_{\bf p}= \sqrt{{\bf p}^2+ m^2}$ the energy of a particle with mass $m$. The creation/annihilation operators follow the anti-commutation relation given by:
\begin{equation}
    \left\{a_{s,{\bf p} }^\dagger, a_{s',{\bf p}'} \right\} = \left\{ b_{s,{\bf p} }^\dagger, b_{s',{\bf p}'} \right\} = \delta_{s s'} \delta_{\bf p p'} ~~. 
\end{equation}
To embrace the spontaneous collapse of the wave function, several nonlinear models have been proposed. In this context,  the Schrodinger equation can be modified as follows:
\begin{eqnarray}\label{eq1}\nonumber
    &&d| \psi_t \rangle = - i\hat{H}  | \psi_t \rangle dt  +\Bigg[ \frac{\sqrt{\gamma}}{m_0}\int d{\bf x} \hat{K}_t({\bf x}) d W_t({\bf x}) + \\\nonumber
    &&-\frac{\gamma}{2m_0^2} \int d {\bf x} d {\bf y} g({\bf x}- {\bf y}) \hat{K}_t({\bf x}) \hat{K}_t({\bf y}) dt \Bigg] | \psi_t \rangle ~~,\\
\end{eqnarray}
with $g({\bf x-y})$ a model-dependent kernel, $m_0$ taken to be of the order of the nucleus mass, $w_t({\bf x}) = \frac{d W_t({\bf x})}{dt}$ a noise field with zero average and correlation equal to  $\mathbb{E} [w_t({\bf x}) w_{t'}({\bf y})]=g({\bf x}-{\bf y}) \delta(t-t')$, and $\hat{K}_t({\bf x}) = \hat{M}({\bf x}) - \langle \psi_t| \hat{M}({\bf x}) | \psi_t \rangle$ with $\hat{M}({\bf x})=  m ~ \hat{N}({\bf x}) $ and where:
\begin{equation}\label{neq}
\hat{N}({\bf x}) = \sum_s \left( a_s^\dagger({\bf x})a_s({\bf x})+b_s^\dagger({\bf x}) b_s({\bf x}) \right).
\end{equation}
Important to highlight that the creation/annihilation operators in Eq. \eqref{neq} are localized in coordinate space. In this work, one also assumes the following structure of the kernel $g({\bf x}-{\bf y})$:
\begin{eqnarray}
~~ g({\bf x}-{\bf y}) =\frac{2 \sqrt2}{\sigma^{3/2}} \times
    e^{-\sigma({\bf x}-{\bf y})^2}~~~, 
\end{eqnarray}
with $\sigma = \frac{1}{4 r^2_C}$.  It can be checked that the coupling constant $\gamma$ has mass dimension $[\gamma]=M^{-2}$. Both $\gamma$ and $r_C$ have bounds due to possible changes in quantum systems \cite{CSLbounds,Bassi13,Bassi17a,Bassi17b,Bassi18}, with estimation given by $\gamma \leq 10^{-34} cm^3 s^{-1}$, and $r_C \approx 10^{-7 }m$ for de CSL model (which leads to  $\gamma \leq 10^{-31} eV^{-2}$, $\sqrt{4\sigma} \approx  eV $, in natural units). Going further, from Eq. \eqref{eq1}, one can integrate out the noise fields and thus obtain the corresponding master equation for the matrix density operator $\hat{\rho}_t$, and on a momentum basis, it is given by;

\begin{eqnarray}\label{eq3}\nonumber
     &&\frac{d\hat{\rho}_t}{dt} =  - {i} [\hat{H}  , \hat{\rho}_t ] +\frac{\gamma m^2}{m_0^2} \int d{\bf p} g({\bf p}) \Big(   \hat{N}({\bf p}) \hat{\rho}_t \hat{N}(-{\bf p}) \\
     &&\hspace{2.cm}-\frac{1}{2}\{ |\hat{N}({\bf p})|^2 , \hat{\rho}_t \} \Big)~~,
\end{eqnarray}
where $\hat{N}({\bf p}) = \sum_{s} \int d{\bf q} \left( a_{s,{\bf p+q} }^\dagger a_{s,{\bf q}}+b_{s,{\bf p+q}}^\dagger b_{s,{\bf q}}\right) $, $d{\bf p} = \frac{d^{D-1} p}{(2 \pi)^{D-1}}$ , and  $g({\bf p})= (2 \pi)^{3/2}e^{-\frac{{\bf p}^2}{4 \sigma }}$. The equation above is a generalization of the GKSL equation, or Lindblad master equation, with Hermitian jump operators $\hat{N}$, and generally it is used to describe the density matrix dynamics for Markovian open systems.
Going further, one can find a formal solution to Eq. \eqref{eq3} and it is the following:
\begin{equation}
\hat{\rho}_t = \frac{e^{i S_{f}\left(\tilde{\Psi},\Psi\right)}}{Z}~~,~~Z = \int \mathcal{D}\left(\tilde{\Psi}, \Psi\right) e^{i S_{f}(\tilde{\Psi},\Psi)} ~~,
\end{equation}
where  $\Psi^{\pm}/\tilde{\Psi}^{\pm}$ are the coherent eigenstates of the creation/annihilation operators,
\begin{eqnarray}\nonumber
    &&\nonumber S_{f}(\Psi^\pm) =\sum_s\int d{\bf p} \int dt \Bigg[ \tilde{\Psi}_s^+(t,{\bf p})( i \partial_t - \epsilon_{\bf p} )\Psi_s^+(t,{\bf p}) +\\\nonumber
    && \hspace{1.cm}- \tilde{\Psi}_s^-(t,{\bf p})( i \partial_t- \epsilon_{\bf p} )\Psi_s^-(t,{\bf p}) +  \\
    &&\hspace{1.cm}+\frac{i \gamma m^2 g({\bf p})}{2 m_0^2} \left|\mathcal{N}^-(t,{\bf p}) - \mathcal{N}^+(t,{\bf p})\right|^2  \Bigg]~~,
\end{eqnarray}
where $\mathcal{N}^\pm(t,{\bf p}) = \sum_s   \int d{\bf q} \tilde{\Psi}^\pm_s (t,{\bf p}+ {\bf q}){\Psi}^\pm_s (t,{\bf q}) $. Important to highlight that the causal structure of the action can be checked via the condition $S_f(\Psi^- =\Psi^+) = 0$. Introducing a new basis, the so-called Keldysh basis, given by:
\begin{equation}\label{kbasis}
    \Psi^{1/2}_s = \frac{1}{\sqrt{2}}\left(\Psi^+_s \pm \Psi^-_s\right)~~,~~  \tilde{\Psi}^{1/2}_s = \frac{1}{\sqrt{2}}\left(\tilde{\Psi}^+_s \mp \tilde{\Psi}^-_s\right)~~,
\end{equation}
the action can be rewritten as:
\begin{eqnarray}\nonumber
    &&S_{f}(\Psi^a) =\sum_s\int d{\bf p} \int dt \Bigg[ \tilde{\Psi}_s^1(t,{\bf p})( i \partial_t - \epsilon_{\bf p} )\Psi_s^1(t,{\bf p}) +\\\nonumber
&&\hspace{1.cm}+\tilde{\Psi}_s^2(t,{\bf p})( i \partial_t - \epsilon_{\bf p} )\Psi_s^2(t,{\bf p})  +\frac{i \gamma m^2 g({\bf p})}{2 m_0^2}\times\\\nonumber
&&\sum_{s'} \int d {\bf q}\int d {\bf q}' \left( \tilde{\Psi}^b_s (t,{\bf p+q})(\hat{\sigma}^{cl})_{bc}{\Psi}^c_s (t,{\bf q})\right) \times \\
&&\hspace{1.cm}\left( \tilde{\Psi}^b_{s'} (t,{\bf q'-p})(\hat{\sigma}^{cl})_{bc}{\Psi}^c_{s'} (t,{\bf q}') \right)   \Bigg]~~.
\end{eqnarray}
where one defines the $2 \times 2$ matrices $\hat{\sigma}^{cl,q}$ as:
\begin{equation}
    (\hat{\sigma}^{cl})_{bc} = 
        \begin{pmatrix}
            1 & 0\\
            0 & 1
        \end{pmatrix}~~,~~ (\hat{\sigma}^{q})_{bc} = 
        \begin{pmatrix}
        0 & 1\\
        1 & 0
        \end{pmatrix} ~~,
\end{equation} 
with $a,b,c = 1,2$.  Now, by writing down the action in terms of four-component Dirac spinors, written as follows:
\begin{eqnarray}\nonumber
    &&\psi^{a}({\bf x}, t) = \sum_{{\bf p},s}  (2 \epsilon_{\bf p})^{-1/2} \times\\
    &&\hspace{.1 cm}\left( \Psi^{a}_{{\bf p},s}(t) u_{{\bf p}, s} e^{ i {\bf p} \cdot {\bf x}}   + (\tilde{\Psi}^c)^{a}_{{\bf p}, s}(t) v_{{\bf p},s} e^{- i {\bf p} \cdot {\bf x}} \right) ~,~
\end{eqnarray}
and defining the Dirac conjugate $\bar{\psi} = \psi^\dagger \gamma^0$ ,

the fermionic action takes the final form given by $S_f = S_{0} + S_{d}$, where:
\begin{eqnarray}\label{sbath2}\nonumber
S_{0}(\bar{\psi},\psi)  = \int dx   \begin{pmatrix}
\bar{\psi}^{1} ~\bar{\psi}^2 
\end{pmatrix} \begin{pmatrix}
i \slashed{\partial}-m   &  0\\
0 & i \slashed{\partial}-m
\end{pmatrix} \begin{pmatrix}
\psi^{1}\\
\psi^2 \end{pmatrix}  ,\\
\end{eqnarray}
with $dx = dt d{\bf x}$, 
 $\slashed{\partial}=\gamma^\mu \partial_\mu = (\gamma^0\partial_t - \gamma^i \cdot \partial^i)$, and the dissipative term reads:
\begin{widetext}
\begin{eqnarray}\nonumber
    &&S_{d} =  \frac{ i  \gamma m^2}{2 m_0^2} \int dx dy dx' dy' \Big( \bar{\psi}_{\alpha}^a(x)   (\hat{\sigma}^{cl})_{a a'}\psi^{a'}_{\alpha'}(x'){\mathcal{Q}}_{\alpha \alpha',\beta \beta'}(x,x',y,y') \bar{\psi}_{\beta}^{b'}(y)   (\hat{\sigma}^{cl})_{b b'}\psi^{b'}_{\beta'}(y') \Big) ~~,\\
\end{eqnarray}
\end{widetext}
with Greek letters representing the spinorial indices, and the kernel ${\mathcal{Q}}_{\alpha \alpha',\beta \beta'}(x,x',y,y')$ is shown in detail in Appendix \ref{appA}. Going further, in the next section, we will look at the Dyson-Schwinger equations of the model.
\section{Dyson-Schwinger equations}
Since the amplification mechanism invalidates perturbative methods, we resort to non-perturbative solutions. One can define the generating functional of the Green functions as follows:
\begin{equation}\label{zfunc}
    Z[\bar{\eta}, \eta] = \int D \bar{\psi} D\psi  e^{i S_f[\bar{\psi},\psi]+i \int dx \left( \bar{\eta}\psi +\bar{\psi}\eta \right)}~~,
\end{equation}
and the generating functional of the connected Green functions can be written as $W = \frac{1}{i}\ln Z$.
Defining the connected Green function as $\mathcal{S}^{bc}_{\beta \zeta}(x,y)=\frac{1}{i}\frac{\delta^2 W}{\delta \eta_\zeta^c(y) \delta \bar{\eta}^b_\beta(x)}\Big{|}_{\eta=\bar{\eta}=0}$ the full fermion propagator. From Eq. \eqref{eom}, one can show that the full fermion propagator respects the following equation:
\begin{eqnarray}\nonumber
   && \int dy\left(  (\mathcal{S}^{-1}_0)^{cb}_{\zeta \beta}(z,y)- \Sigma^{cb}_{\zeta \beta}(z,y)  \right) \mathcal{S}_{\beta \delta}^{bd}(y,w)  =\\
   &&\hspace{3.cm}= i \delta^{cd} \delta_{\delta \zeta}\delta(z-w)~~,
\end{eqnarray}
\\
where the self-energy $\Sigma_{ab}(x,y)$, in the 1-loop approximation, is shown in Appendix \ref{AppB}. Now, decomposing the full fermion propagator as $\mathcal{S} = \begin{pmatrix}
        S^R & S^K\\
        0 & S^A
    \end{pmatrix}$, one has:
    
\begin{equation}\label{DS0}
    \begin{pmatrix}
         i\slashed{\partial}-m  - \Sigma^R& - \Sigma^K\\
       0 & i\slashed{\partial}-m - \Sigma^A
    \end{pmatrix} \circ \begin{pmatrix}
        S^R & S^K\\
        0 & S^A
    \end{pmatrix} = \begin{pmatrix}
        i & 0\\
        0 & i
    \end{pmatrix}~~,
\end{equation}
where the ``$\circ$" symbol means the convolution operation in coordinate space. Looking at the components of Eq. \eqref{DS0}, one can obtain the Dyson-Schwinger (DS) equations as follows: 
\begin{equation}\label{DS1}
(  i\slashed{\partial}-m - \Sigma^{R/A})~\circ~ S^{R/A} =i ~~.
\end{equation}

Going further, without loss of generality one can assume $S^K = S^R \circ F - F \circ S^A$, with $F(x,y)$ an spinorial function of $x$ and $y$, and after some manipulations one finds:
\begin{widetext}
\begin{eqnarray}\label{Feq}
    &&(\overrightarrow{\slashed{\partial}}_x - m) \circ F(x,y) - F(x,y)o(\overleftarrow{\slashed{\partial}}_y - m) =- \Sigma^K(x,y) + (\Sigma^R \circ F- F \circ \Sigma^A)(x,y)~~.
\end{eqnarray}
\end{widetext}
The Eq. \eqref{Feq} is the generalization of the Boltzmann equation.  Now, if one looks for translationally invariant solutions, 
one can show that $ \Sigma^{cb}(q)$ respects the following identity:
\begin{eqnarray}\label{sigmaEq}\nonumber
  &&U({\bf p})  \Sigma^{cb}(p)U^\dagger({\bf p}) =\\\nonumber
  &&=\frac{\gamma m^2}{m_0^2}  \Big[ \int dp' g({\bf p-p'})U({\bf p'})\mathcal{S}^{c b}(p')U^\dagger({\bf p}') \Big]~,~\\
\end{eqnarray}
with $U({\bf p})$ the Foldy-Wouthuysen (FD) unitary matrix shown in Appendix \ref{appA}. Based on the structure of the kernel $g({\bf p})$, one can affirm that the self-energy is a function of ${\bf p}$. One can also show that, by use of the causal structure of $\mathcal{S}^{cb}(p)$, that: 
\begin{equation}\label{Recond}
    Re[\Sigma^R(p)] =  \frac{1}{2}\delta^{cb}\Sigma^{cb}(p) = 0~~,
\end{equation}
which implies that the self-energy has only imaginary non-trivial solutions. Therefore, one defines $\Gamma(p)=-Im[\Sigma^R(p)]$, and the gap equations reads:
\begin{eqnarray}\label{sigmaEq2}\nonumber
    &&U({\bf p})\Gamma({\bf p})U^\dagger({\bf p}) = \\
    &&-\frac{\gamma m^2}{m_0^2}  \Big[ \int dp' g({\bf p-p'})U({\bf p'})\Delta(p')U^\dagger({\bf p}') \Big]~,~
\end{eqnarray}
with $\Delta(p) = i\mathcal{S}^R(p)-i\mathcal{S}^A(p)$ the so-called spectral function. 
Assuming the structure of the inverse of the retarded full fermionic propagator given by $(\mathcal{S}^R)^{-1}(p)= \mathcal{A}^{-1}(p)\left( \gamma_0 p_0 - \mathcal{Z}(p) \gamma \cdot {\bf p} - B(p) \right) $, where $\mathcal{A}(p)$ and $\mathcal{Z}(p)$ are the fermion wavefunction renormalization functions, and $B(p)$ is the dynamical fermion mass, and both are complex functions of $p$. Thus, due to the condition shown in Eq. \eqref{Recond}, one can affirm that $ \mathcal{A}(p) = 1$, and therefore:
\begin{equation}\label{wfren}
\mathcal{Z}({\bf p}) = 1 + \frac{i}{{\bf p}^2}Tr[\gamma \cdot {\bf p} \Gamma({\bf p})] ~~,~~ 
\end{equation}
and
\begin{equation}
    B({\bf p}) = m -i Tr[  \Gamma({\bf p})]~~,
\end{equation}

Using the trace property of cyclicity, one can get rid of the complicated structure of the FW matrices, and the DS equation reduces to: 
\begin{equation}\label{GAP1}
    \mathcal{Z}({\bf p}) = 1 -\frac{i}{{\bf p}^2}\frac{\gamma m^2}{m_0^2}  \int dp' g({\bf p-p'})Tr[\gamma \cdot {\bf p} \Delta(p') ] ~~,~~ 
\end{equation}
\begin{equation}\label{GAP2}
B({\bf p}) = m   -i\frac{\gamma m^2}{m_0^2}  \int dp' g({\bf p-p'})Tr[\Delta(p') ]~~,
\end{equation}

\subsection{Boltzmann equation}
In Appendix \ref{AppC}, one shows in detail that the Keldysh component of the fermionic propagator is the object one need to look to find the system's information about the colective properties, in such a way that Assuming the Wigner function $F(x,y)$ a scalar in spinor space, the system obeys the following Boltzmann equation (see Appendix \ref{AppC} for details):
\begin{equation}
    \gamma^\mu    \frac{\partial F(X,q)}{\partial X^\mu} = - i ~U^\dagger({\bf p})\mathcal{I}(X,q) U({\bf p})~~,
\end{equation}
where, in the first order in gradient expansion, one has:
\begin{equation}
    \mathcal{I}(X,q) = i \frac{\gamma m^2}{ m_0^2}  \int dq' g({\bf q-q'})\left( F(X,q')- F(X,q)\right)\mathcal{D}(q')~~,
\end{equation}

with $\mathcal{D}(q) = U({\bf q}) \Delta(q) U^\dagger({\bf q})$ and where $\Delta(q) = i\mathcal{S}^R(q)-i\mathcal{S}^A(q)$ the spectral function. It is easy to show that a conserved current $J^\mu$ can be written as follows:
\begin{eqnarray}\nonumber
    J^\mu (X) =  \int d q  F(X,q) Tr[\gamma^\mu\Delta(q)]~,~ \frac{\partial J^\mu (X) }{\partial X^\mu} = 0~.~\\
\end{eqnarray}

Thus, one can interpret the currents $J^\mu$ as the electromagnetic current. Going further, the quantity is defined as:
\begin{eqnarray}
    T^{\mu \nu} (X) =   \int d q  q^\nu  F(X,q) Tr[\gamma^\mu \Delta(q)]~~,~~ 
\end{eqnarray}
is not conserved, and its divergence $\frac{\partial T^{\mu \nu}  }{\partial X^\nu}$ can be written as:
\begin{eqnarray}\nonumber
   &&\frac{\partial T^{\mu \nu}  (X)}{\partial X^\nu} =Q^\mu(X)= \\\nonumber
   &&=-i \frac{\gamma m^2}{ m_0^2} \int dq  \int dq'\Big(  {q}^\mu  g({\bf q-q'})\left( F(X,q')- F(X,q)\right)\times \\
   && \hspace{4.cm}Tr[\Delta(q') \Delta(q)] \Big)  ~~.~~
\end{eqnarray}
Here, $T^{\mu \nu}(X)$ can be interpreted as an energy-momentum tensor, and $Q^\mu(X)$ as the heat flow due to the spontaneous collapse of the quantum system. 

\subsection{Equilibrium symmetry}
In the context of Keldysh field theory, the thermal equilibrium state is given by the state that respects a specific symmetry. For fermionic fields, the symmetry is given by an antiunitary time reversal with an additional $\mathbb{Z}_2$ transform \cite{Aron18,Sieberer25,Altland21}, i.e.:
\begin{equation}
  \mathcal{T}_{\beta} : \begin{cases} 
   \psi^{\pm}(t) \to \psi^{\pm}_{\beta}(t)= \mp (\tilde{\psi}^\pm)(- t \pm i \beta/2)\\
   \tilde{\psi}^{\pm}(t) \to \tilde{\psi}^{\pm}_{\beta}(t)= \pm ({\psi}^\pm)(- t \pm i \beta/2)
   \end{cases}
\end{equation}
where $\beta^{-1}$ is the equilibrium temperature. On the Keldysh basis and in momentum space, one can write:
\begin{eqnarray}\nonumber
   &&\mathcal{T}_{\beta}:
   \psi^{a}(p_0,{\bf p})  \to \psi^a_{\beta}(p_0,{\bf p})= \\\nonumber
   &&=\left(\sinh(p_0 \beta/2) (\hat{\sigma}^{cl}_{ab}) + \cosh(p_0 \beta/2) (\hat{\sigma}^{q}_{ab})\right)  (\tilde{\psi}^{b})(-p_0,{\bf p}) ~~,\\
\end{eqnarray}
\begin{eqnarray}\nonumber
   &&\mathcal{T}_{\beta}:
   \tilde{\psi}^{a}(p_0,{\bf p})  \to \tilde{\psi}^a_{\beta}(p_0,{\bf p})= \\\nonumber
   &&=-\left(\sinh(p_0 \beta/2) (\hat{\sigma}^{cl}_{ab}) + \cosh(p_0 \beta/2) (\hat{\sigma}^{q}_{ab})\right)  ({\psi}^{b})(-p_0,{\bf p}) ~~.\\
\end{eqnarray}
For Lindbladian systems, the equilibrium symmetry imposes strong restrictions on the Green functions.   Particularly, the equilibrium symmetry imposes that the two-point functions must respect the following identity \cite{Sieberer25}:
\begin{equation}
    \mathcal{S}^K_{eq}(p_0,{\bf p}) = \tanh(\beta p_0/2) \left( \mathcal{S}^R(p_0,{\bf p}) - \mathcal{S}^A(p_0,{\bf p}) \right) ~~.
\end{equation}
The above relation is the well-known fluctuation-dissipation relation (FDR). Going further, the bilinear object that composes the quartic vertex is clearly not invariant under $\mathcal{T}_{\beta}$. Thus, the vertex and, in consequence, the total action will be invariant under thermal symmetry only in the limit $\beta \to 0$. Thus, one can affirm that the CSL model is an out-of-equilibrium system with no steady states.  

\section{results}
We now present both perturbative and non-perturbative solutions to the gap equations, highlighting their impact on particle decay rates. For instance, by using the values of the constants  $\gamma \approx 3.2 \times 10^{-31} eV^{-2}$, $m_0 \approx 1 GeV$ and $\sqrt{4\sigma} \approx  1 eV $ given by the GRW model \cite{GRWref}, due to the amplification mechanism which objectively appears on the dependence on the values of mass $m$ of the particle on the effective coupling $\tilde{\gamma}=\gamma m^2/m_0^2 \approx 10^{-49} eV^{-4} m^2 $, and also the characteristics which the effective coupling constant has dimension of $(mass)^{-2}$, one assumes that a perturbative expansion is not applicable. Thus, one will look into the non-perturbative approach to extract information. Independent of the strength of the effective coupling constant, due to its imaginary character, the effective mass becomes complex.  This new feature, which arose as a result of the spontaneous collapse, will be analyzed in the sequel.

\subsection{Non-perturbative aspects}
\begin{figure}[H]
    \centering
    \includegraphics[width=1\linewidth]{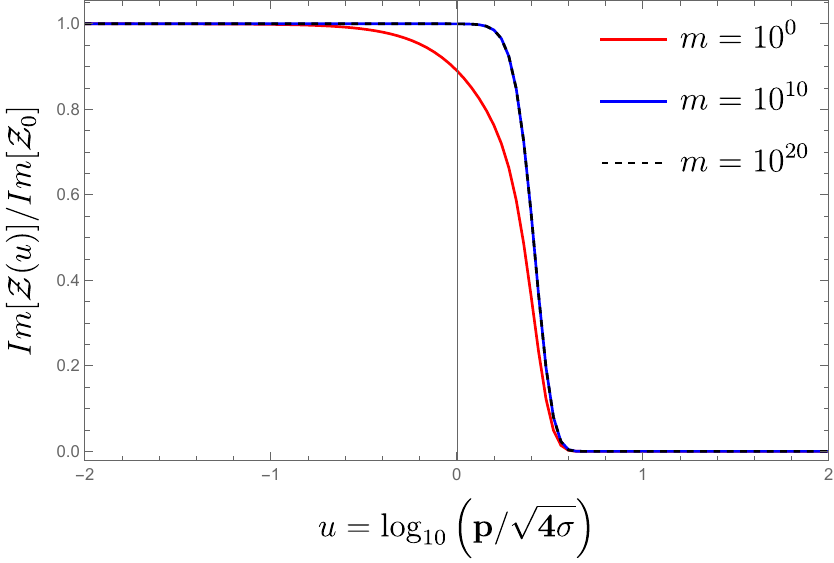}
    \caption{Plot of the solutions of the non-perturbative gap equations for $Im[\mathcal{Z}(u)]$ , normalized with respect to $ \mathcal{Z}_0=\lim_{{\bf p \to 0}}\mathcal{Z}({\bf p})$,  with $u= \log_{10}|{\bf p}|/\sqrt{4\sigma}$. }
    \label{fig1}
\end{figure}

\begin{figure}[H]
    \centering
    \includegraphics[width=1\linewidth]{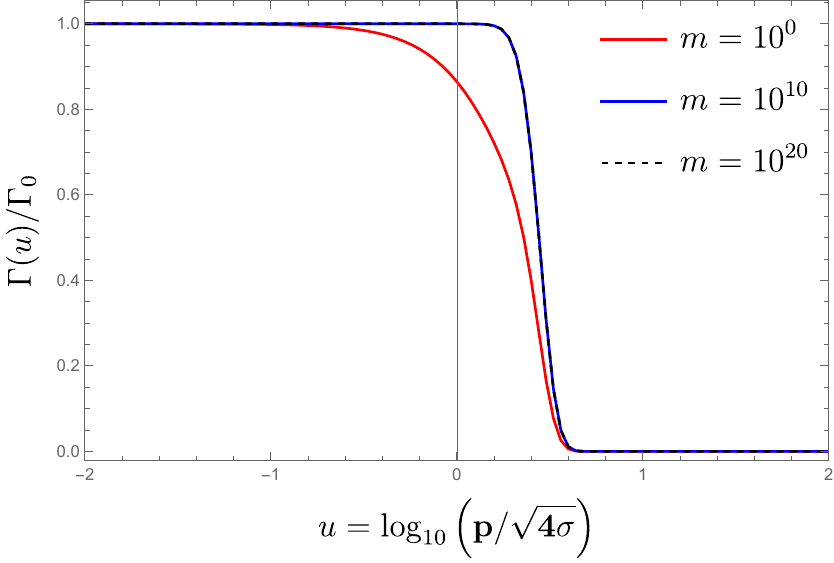}
    \caption{Plot of the solutions of the non-perturbative gap equations for $\Gamma(u)=Im[M(u)]$, normalized with respect to $ \Gamma_0=-\lim_{{\bf p \to 0}}Im[M({\bf p})]$,  with $u= \log_{10}|{\bf p}|/\sqrt{4\sigma}$. }
    \label{fig2}
\end{figure}

\begin{figure}[H]
    \centering
\includegraphics[width=1\linewidth]{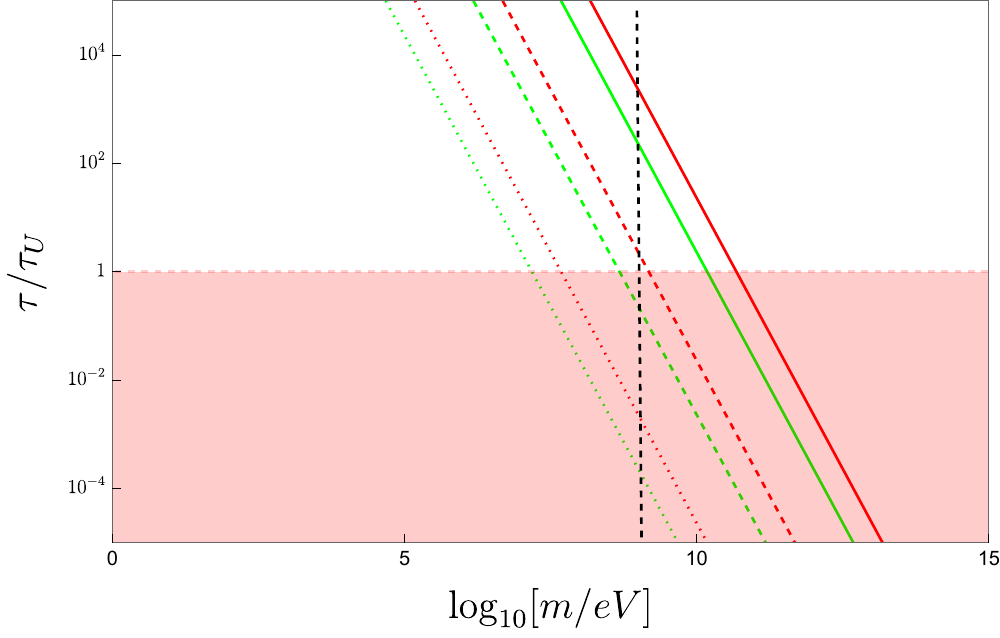}
\caption{Non-perturbative solutions of the mean lifetime fraction  $\tau/\tau_U$, with $\tau=1/\Gamma_0$ and $\tau_U =13.8 \times 10^9 $years, as a function of $\log_{10}[m/eV]$. The solid green, dashed green, and dotted green lines represent the non-perturbative solutions for $\lambda= 10^{-17} s^{-1}$ and $r_C =10^{-7}m$, $r_C =10^{-6} m$, $r_C =10^{-5} m$, respectively. The solid red, dashed red, and dotted red lines represent the non-perturbative solutions for $\lambda= 10^{-18} s^{-1}$ and $r_C =10^{-7}m$, $r_C =10^{-6} m$, $r_C =10^{-5} m$, respectively. The black dashed line represents the proton ($m_p=928$ MeV) and the red region represents the instability region. }
    \label{fig3}
\end{figure}

Without expanding over $\tilde{\gamma}$, one needs to solve the non-linear Fredholm-like integral equations given by Eqs. \eqref{GAP1} and \eqref{GAP2}. By use of numerical techniques, one can find the solutions for a given $\tilde{\gamma}$, and the solutions for $\mathcal{Z}({\bf q})$ and $\Gamma({\bf q})$ are shown in Figs. \ref{fig2}, \ref{fig3}  and \ref{fig1}. 

\begin{figure}[H]
    \centering
    \includegraphics[width=1\linewidth]{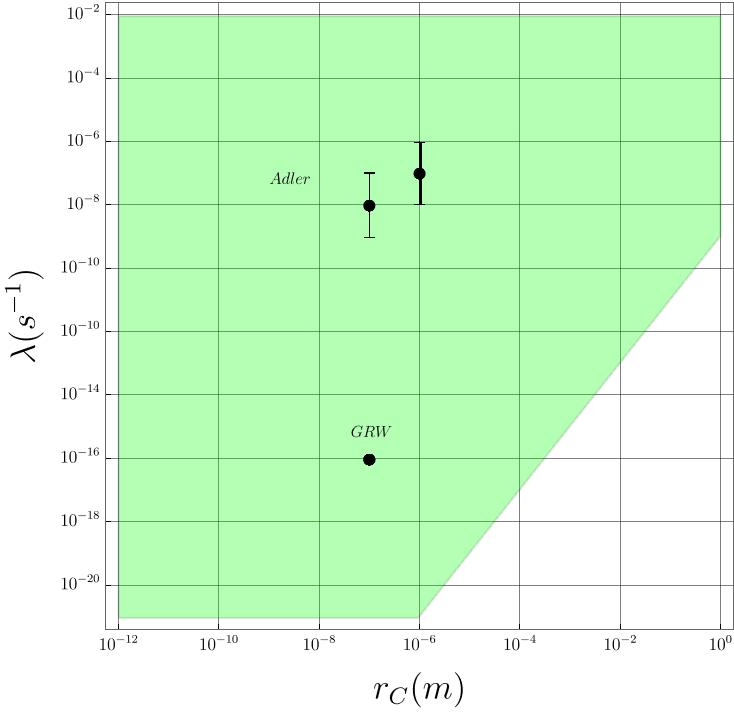}
    \caption{Parameter space plot for the SC model.  The light green exclusion region comes from the proton stability in the non-perturbative approach. For reference, one includes the GRW \cite{GRWref} values ($\lambda = 10^{-16} s^{-1}$, $r_C  = 10^{-7} m$ ) and the values proposed by Adler \cite{Adlerref} ($\lambda = 10^{-8\pm 2} s^{-1}$, $r_C = 10^{-7} m$ ) and ($\lambda = 10^{-6\pm 2} s^{-1}$, $r_c = 10^{-6} m$). }
    \label{fig4}
\end{figure}

Assuming the mean lifetime as $\tau = 1/\Gamma_0$, with $\Gamma_0= \lim_{\bf p \to 0} \Gamma({\bf p})$, one can estimate the mean lifetime of massive fermions due to spontaneous collapse. Assuming some representative values of the constants $\lambda= \gamma/(2 \pi r_c^2)^{3/2}$ and $r_c$, one can plot the mean lifetime as a function of the fermionic particle mass, and one shows it in Fig. \ref{fig3}. For instance, the proton with mass $m_p \approx 938 MeV$ must be stable within our universe's age ($\tau_U \approx 13.8 \times 10^9 $ years), and thus its mean lifetime must be greater than $\tau_U$. By taking advantage of the gaussian profile of the kernel $g({\bf p})$ , the gap equation for the function $B(x)$, assuming $m\gg \sqrt{4 \sigma}$ can be approximate written as:
\begin{eqnarray}\nonumber
&&B(x) = m   -2i\frac{\gamma m^2}{m_0^2} (4\sigma)^{3/2} \times \\\nonumber
&&Re \int_0^\infty dy \mathcal{G}(x,y)\frac{y^2 B( y)/(\sqrt{4\sigma})}{ \sqrt{\mathcal{Z}(y)^2 y^2 + B(y)^2/(4 \sigma)}}=\\
&&\approx m   -2i\frac{\gamma m^2}{m_0^2} (4\sigma)^{3/2} Re \int_0^\infty dy \mathcal{G}(x,y)y^2 ~~.
\end{eqnarray}
and therefore, for $m \gg \sqrt{4 \sigma}$ the imaginary part can be written as $\Gamma_0 \approx \frac{3 }{2 \sqrt{2}} \frac{\gamma m^2}{m_0^2} (4\sigma)^{3/2}$. From this result, one can restrict the parameter space by imposing proton stability, and the resulting allowed parameter space is shown in Fig. \ref{fig4}.

\section{final remarks}
In this work, we analyze the effect of the spontaneous collapse hypothesis on the dynamics of a relativistic fermionic system. Using Keldysh formalism, we construct an action for the system, and via the Dyson-Schwinger equations, we can analyze the dynamical symmetry breaking with a truncation at 1-loop order.  The result shows a qualitative change across the scale determined by $\sqrt{4\sigma}$, i.e., the energy scale where the collapse effects are non-negligible,  the self-energy solutions change qualitatively. This behavior is a QFT manifestation of the quantum-to-classical transition.  Our result shows a new mechanism that introduces a dynamical generation of fermionic complex masses. Interestingly, the non-perturbative solutions of the SC model show that the spontaneous collapse of the wave-function can generate measurable effects on single particles. One also finds that $\Gamma({\bf p})<0$, which ensures the stability of the system through perturbations. Due to the Gaussian profile of $g({\bf p})$, one finds that the solutions shown in figs. \ref{fig2} and \ref{fig3} are finite in the infrared (IR) region, but vanish in the ultraviolet (UV) region, with the transition region determined by $\sqrt{4 \sigma}$. Therefore, one can affirm that the SC affects only the IR of the relativistic fermionic system, and its results impose bounds on the coupling constants $\gamma$ and $r_C$. Assuming the stability of the proton over the universe's age, one shows a phase portrait of $\lambda \times r_c$. By doing so, one generates new bounds on its parameters, as shown in Fig. \ref{fig4}. Our result, based on proton stability, excludes the parameter values for the GRW model and further restricts the parameter space. 

Another novelty resides in the fact that the results do not respect Lorentz invariance, a feature that is explicitly shown in the self-energy, which depends on $|\bf p|$ only. This fact is expected since, in general, Lindblad's actions lack Lorentz invariance \cite{lindref3}, \cite{lindref4}, \cite{gomes24}.
The analysis also reveals that the violation of Lorentz invariance lies in the infrared (IR) regime, where collapse effects become non-negligible. In contrast, Lorentz symmetry is restored in the ultraviolet (UV) limit, consistent with the Gaussian profile of the collapse kernel $g({\bf p})$. This behavior suggests that SC models may provide an effective description of the quantum-to-classical transition in the IR, without conflicting with relativistic symmetries at high energies.

Important to highlight that the gap equation and the respective complex masses do not involve the Keldysh component of the fermionic propagator at 1-loop. This fact is due to the structure of the self-interaction, which relates only to interactions between the classical scalar current component $m \bar{\psi} \hat{\sigma}^{cl}\psi$. The physical implication is that the results do not depend on the thermodynamics of the fermionic system. One has also checked that the action is not invariant under thermal symmetry $\mathcal{T}_{\beta}$, and therefore, its solutions are non-equilibrium solutions and cannot be rewritten as a subsystem coupled to a bath. Besides a strong bound due to the causality properties of the Keldysh action, an increased analysis of an N-loop truncation \cite{2piref} can improve these results. 

Going further, the collective behavior of the many-body fermionic system can be settled by the hydrodynamics of the system, and it can be obtained through the Keldysh component of the propagator. By use of this analysis, a conserved charge is found, whereas the energy-momentum tensor is shown to be non-conserving, consistent with other results about CSL models, in which the well-known property that the noise generates a small but not negligible heating effect \cite{Bassi13},\cite{diss23}. One way to circumvent this issue is to introduce dissipative terms on the jump operators $\hat{L}$, as $\nabla_{\bf x}\cdot J({\bf x})$, with $J({\bf x}) \propto \left( a^\dagger({\bf x}) \nabla_{\bf x} a({\bf x}) - h.c.\right)$ such as proposed on Ref. \cite{diss23}. The introduction of these new kinds of jump operators can bring the system to thermal equilibrium, and should introduce new terms on the Boltzmann equation shown in Eq. \eqref{kinetic}. However, the Lorentz violation can be intensified by these dissipative factors.

Another important insight is that the Keldysh formalism is compatible with Lorentz invariance, in opposition to the Lindblad equation \cite{lindref3,lindref4,gomes24}. This new procedure of analyzing the collapse problem can be an opportunity to describe the collapse of the wave function in a relativistic manner, by defining proper Lorentz-invariant currents in which the noise field couples. These and other modifications will be the target of investigation in further work.

\section*{Acknowledgments}
YMPG thanks MFSA and NPN for the major revision of the manuscript. YMPG is supported by a postdoctoral grant from Funda\c{c}\~ao Carlos Chagas Filho de Amparo \`a Pesquisa do Estado do Rio de Janeiro (FAPERJ), Grant No. E-26/200427/2025. 
%


%
\begin{widetext}
\appendix 
\section{Spinorial definitions}\label{appA}
This appendix shows the orthonormalized Dirac eigenvectors used in the article. Starting from the Dirac equation in four dimensions, one has:
\begin{equation}
    (i \gamma^\mu \partial_\mu - m ) \psi(x) = 0~~,
\end{equation}
with $\psi(x)$ a generic four-component spinor and $\gamma^\mu$, $\mu=0,i$ ($ i=x,y,z$), the $4 \times 4$ Dirac matrices in the Dirac representation can be written as:
\begin{equation}
   \gamma^0 = \begin{pmatrix}
        \mathds{1}_{2 \times 2} & 0 \\
        0 & -\mathds{1}_{2 \times 2} 
    \end{pmatrix} ~~,~~ \gamma^i = \begin{pmatrix}
        0 & \sigma^i \\
        - \sigma^i & 0
    \end{pmatrix}~~,
\end{equation}
where $\sigma^i$ are the Pauli matrices.
By use of the Foldy-Wouthuysen unitary matrix given by $U({\bf p})$ \cite{FW}, one can write down explicitly the eigenspinors as follows:
\begin{eqnarray}
    &&u_{{\bf p},s} = U({\bf p})u_{0,s}  ~~,~~ v_{{\bf p},s} = U({\bf p}) v_{0,s}~~,
\end{eqnarray}
 
with
\begin{equation}
    u_{0,s} = N_{0}\begin{pmatrix}
    \xi_s\\
    0 \\
    \end{pmatrix}~~,~~
            v_{0,s} = N_{0}\begin{pmatrix}
    0\\
    \eta_s \\
            \end{pmatrix}~~,
\end{equation}
where $\xi_s$, $\eta_s$ are orthonormal bispinors, $N_{0}=\sqrt{2\epsilon_{\bf p}}$, and:

\begin{equation}
    U({\bf p}) = e^{\gamma \cdot {\bf \hat{p}} \theta_{\bf p}}=\frac{\left( \epsilon_{\bf p} + m + \gamma \cdot {\bf p} \right)}{\sqrt{2 \epsilon_{\bf p} (\epsilon_{\bf p} +m)} }~~,
\end{equation}
with $\tan 2\theta_{\bf p} = |{\bf p}|/m$, $U(-{\bf p}) = U^\dagger({\bf p}) = U^{-1}({\bf p})$, and  $\gamma_0 U = U^\dagger \gamma_0$. The eigenvectors respect the following identities:
\begin{equation}
\bar{u}_{{\bf p}, s} u_{{\bf p}, s'} =2 m \delta_{s s'}~~,~~ \bar{v}_{{\bf p}, s} v_{{\bf p}, s'} =-2 m \delta_{s s'}~~,
\end{equation}
and
\begin{equation}
\bar{u}_{{\bf p}, s} \gamma^\mu u_{{\bf p}, s'}=\bar{v}_{{\bf p}, s} \gamma^\mu v_{{\bf p}, s'} =2 \hat{p}^\mu \delta_{s s'}~~.
\end{equation}

Going further, in a compact notation one can define a set of eigenvectors as $W_r({\bf p}) $ with $r=1,2,3,4$, i.e., $W_r({\bf p})=  (2 \epsilon_{\bf p})^{-1/2} u_{s,{\bf p}}$ if $r=1,2$, and $ W_r({\bf p})= (2 \epsilon_{\bf p})^{-1/2} v_{s,-{\bf p}} $, if $r=3,4$. In this notation, one has:
\begin{equation}
\bar{W}_r({\bf p}) {W}_{r'}({\bf p}) = \eta_r \frac{m}{\epsilon_{\bf p}} \delta_{r r'} ~~,~~
    \bar{W}_r({\bf p}) \gamma^\mu {W}_{r'}({\bf p}) = \frac{\hat{p}^\mu}{\epsilon_{\bf p}} \delta_{r r'} ~~,
\end{equation}
with $\eta_r= + 1(-1)$ for $r=1,2(r=3,4)$. 
Also, the following properties are satisfied: 
\begin{eqnarray}\nonumber
&&\sum_r {W}_{r}({\bf p}) \bar{W}_{r}({\bf p'})= U({\bf p})\left(  \sum_r {W}_{r}(0) {W}^{\dagger}_{r}(0) \right) U^\dagger({\bf p'}) \gamma_0 =\\
&&= U({\bf p}) U^\dagger({\bf p'}) \gamma_0 ~~,
\end{eqnarray}
and
\begin{eqnarray}\nonumber
&&\sum_r \eta_r{W}_{r}({\bf p}) \bar{W}_{r}({\bf p'})= U({\bf p})\left(  \sum_r \eta_r{W}_{r}(0) {W}^{\dagger}_{r}(0) \right) U^\dagger({\bf p'}) \gamma_0 =\\
&&= U({\bf p}) \gamma_0 U^\dagger({\bf p'}) \gamma_0 =  U({\bf p})U({\bf p'})~~.
\end{eqnarray}
Lastly, one can define the particle/antiparticle projectors as follows:
\begin{equation}
    \mathcal{P}_{\pm}({\bf p}) = U({\bf p}) \frac{(1\pm \gamma_0)}{2} U^\dagger({\bf p})= \frac{1}{2} \left(1 \pm  U^2({\bf p})\gamma_0 \right) = \frac{1}{2} \left(1 \pm \frac{(m + \gamma \cdot{\bf p})}{\epsilon_{\bf p}} \gamma_0\right) ~~,
\end{equation}
which obeys $\mathcal{P}_{+} + \mathcal{P}_{-} = 1$, and $\mathcal{P}_{+}\mathcal{P}_{-}=\mathcal{P}_{-}\mathcal{P}_{+}=0$. One also has $\bar{\mathcal{P}}_{\pm} = \gamma_0 \mathcal{P}^\dagger_{\pm} \gamma_0 = \mathcal{P}_{\pm}$. By use of the particle/antiparticle projectors, the FW matrix can be rewritten as follows:
\begin{equation}
   U({\bf p})=\frac{\sqrt{2\epsilon_{\bf p}}}{\sqrt{(\epsilon_{\bf p} +m)} } \left( \mathcal{P}_{+}({\bf p})\frac{(1+\gamma_0)}{2}+ \mathcal{P}_{-}({\bf p}) \frac{(1-\gamma_0)}{2} \right)~~.
\end{equation}


\section{Dyson-Schwinger equation}\label{AppB}

The action of the model is $S_f = S_{0,f}  + S_{d}$, where:
\begin{eqnarray}\label{sbath2}
S_{f,0}(\bar{\psi},\psi) = \int dx dy  \bar{\psi}^{a}_\alpha(x) (\mathcal{S}_0^{-1})_{\alpha \beta}^{ab}(x,y) \psi^b_{\beta} (y)~~,
\end{eqnarray}
with $(\mathcal{S}_0^{-1})_{\alpha \beta}^{ab}(x,y)= -(\hat{\sigma}^{cl})^{ab} (i\slashed{\partial}-m)_{\alpha \beta} \delta(x-y)$. The dissipative part of the action reads:
\begin{eqnarray}\nonumber
    &&S_{d} =  \frac{ i  \gamma m^2}{2 m_0^2} \int dx dy x' dy' \Big( \bar{\psi}_{\alpha}^a(x)\hat{\sigma}^{cl}_{ab} \psi^{b}_{\beta}(y){\mathcal{Q}}_{\alpha \beta,\alpha' \beta'}(x,y,x',y') \bar{\psi}_{\alpha'}^{a'}(x')\hat{\sigma}^{cl}_{a'b'} \psi^{b'}_{\beta'}(y') \Big) ~~,
\end{eqnarray}
where, in momentum space:
\begin{eqnarray}
   &&{\mathcal{Q}}_{\alpha \beta,\alpha'\beta'}(q,p,q',p') =   \Lambda_{\alpha \beta}({\bf q},{\bf p})  \Lambda_{\alpha' \beta'}({\bf q'},{\bf p'}) \delta(p+p'+q+q') ~~,
\end{eqnarray}
with
\begin{equation}
\Lambda_{\alpha \alpha'}({\bf p},{\bf q}) = \sqrt{g({\bf p+q})} \sum_{r} \left[ \eta_r  \gamma_0 W_r({\bf p})\bar{W}_r({\bf q})\gamma_0\right]_{\alpha \alpha'} = \sqrt{g({\bf p+q})} \left[ U^\dagger({\bf p}) U^\dagger({\bf q})\right]_{\alpha \alpha'} ~~,
\end{equation}
 
where $W_r$ and $U$ are the same objects defined in the previous appendix. Going further, the generating functional can be written as follows:
\begin{equation}\label{zfunc}
    Z[\bar{\eta}, \eta] = \int D \bar{\psi} D\psi  e^{i S_f[\bar{\psi},\psi]+i \int dx \left( \bar{\eta}\psi +\bar{\psi}\eta \right)}~~.
\end{equation}

In terms of the connected generating function $W= \frac{1}{i} \ln Z$, and apllying $\frac{\delta }{\delta \eta^\delta_d(w)}$ at $\bar{\eta}=\eta=0$ one finally reaches the DS equation:
\begin{eqnarray}\label{eom}\nonumber
&&  \int dy (\mathcal{S}^{-1}_0)^{cb}_{\zeta \beta}(z,y) \mathcal{S}_{\beta \delta}^{bd}(y,w) + \\\nonumber
&& + \frac{i \gamma m^2}{2 m_0^2}  \int dy dx' dy' \Big[ \mathcal{Q}_{\zeta  \beta, \alpha' \beta'}(z,x',y,y') \delta^{cb} \delta^{a' b'} - \mathcal{Q}_{\alpha'  \beta, \zeta \beta'}(y,x',z,y')\delta^{a'b}\delta^{c b'} \Big] \times\\\nonumber
&& \Bigg[\mathcal{S}_{\beta' \alpha'}^{ b' a'}(y',y) \mathcal{S}_{\beta \delta}^{bd}(x', w) - \mathcal{S}_{\beta \alpha'}^{b a'}(x',y) \mathcal{S}_{\beta' \delta}^{b' d}(y' ,w) - (W^{(4)})_{\beta' \alpha' \beta \delta}^{b' a' b d}(y',y,x',w)\Bigg] = i \delta^{dc} \delta_{\delta \zeta}\delta(z-w)  \nonumber\\
\end{eqnarray}
with $\mathcal{S}^{cb}_{\beta \alpha}(y,x)=\frac{1}{i}\frac{\delta^2 W}{\delta \eta^b_{\alpha}(x) \delta \bar{\eta}^c_{\beta}(y)}\Big{|}_{\eta=\bar{\eta}=0}$  the connected 2-point Green function  and 
\begin{equation}
 (W^{(4)})_{\beta' \alpha' \beta \delta}^{b' a' b d}(y',y,x',w) = \frac{\delta^4 W}{\delta \eta^d_{\delta}(w) \delta \bar{\eta}^b_{\beta}(x') \delta {\eta}^{a'}_{\alpha'}(y) \delta \bar{\eta}^{b'}_{\beta'}(y')}\Bigg{|}_{\eta=\bar{\eta}=0}  ~~.
\end{equation}
with $ W^{(4)}_{abcd}(y,x,z,w)$ the connected 4-point Green function. Thus, the self-energy $\Sigma_{ab}(x,y)$, in the 1-loop approximation, is given by:

\begin{eqnarray}\label{self}\nonumber
  && \Sigma^{cb}_{\zeta \beta}(z,y) = -\frac{\gamma m^2}{2 m_0^2}  \delta^{cb} \int  dx' dy' \Bigg[ {\mathcal{Q}}_{\zeta \beta, \alpha' \beta'}(z,y,y',x')+{\mathcal{Q}}_{\alpha' \beta',\zeta \beta }(x',y',z,y) \Bigg]\mathcal{S}^{b' b'}_{\beta '\alpha'}(x',y') +\\
   &&\hspace{1.cm}+ \frac{\gamma m^2}{2 m_0^2} \int  dx' dy' \Bigg[ {\mathcal{Q}}_{\alpha' \beta, \zeta \beta'}(x',y,z,y')+{\mathcal{Q}}_{\zeta \beta' ,\alpha' \beta}(z,y',x',y)\Bigg]\mathcal{S}^{c b}_{\beta '\alpha'}(y',x')~~.
\end{eqnarray}
Finally, in momentum space, one finds:
\begin{eqnarray}\nonumber
    &&\Sigma^{cb}_{\zeta \beta}(p) = -\frac{ \gamma m^2}{m_0^2} \delta^{cb} \Lambda_{\zeta \beta} ({\bf p},-{\bf p}) \int dp' \Lambda_{\alpha' \beta'} ({\bf p'},-{\bf p'})\mathcal{S}^{b'b'}_{\beta' \alpha'}(p')+\\
    && +\frac{ \gamma m^2}{m_0^2}  \int dp' \Lambda_{\zeta \beta'} ({\bf p},-{\bf p'})\mathcal{S}^{c b}_{\beta' \alpha'}(p')\Lambda_{\alpha' \beta} ({\bf p}',-{\bf p})~~,
\end{eqnarray}
and using the causality condition given by $\int dp_0 S^{b'b'}(p) =\int dp_0 \left( \mathcal{S}^{R}(p) + \mathcal{S}^A(p)\right) =0 $, one finds:
\begin{eqnarray}
    &&\Sigma^{cb}(p) = \frac{ \gamma m^2}{m_0^2} U^\dagger({\bf p}) \Big[ \int dq' g({\bf p-p'})U({\bf p'})\mathcal{S}^{c b}(p')U^\dagger({\bf p}') \Big]U({\bf p})~~.
\end{eqnarray}
\section{ Boltzmann equation}\label{AppC}
By using the equation $S^K(x,y) = \left( S^R \circ F - F \circ S^A\right) (x,y)$, with $F(x,y)$ one finds:
\begin{eqnarray}
    &&(\overrightarrow{\slashed{\partial}}_x - m) \circ F(x,y) - F(x,y)\circ(\overleftarrow{\slashed{\partial}}_y - m) =- \Sigma^K(x,y) + (\Sigma^R \circ F- F \circ \Sigma^A)(x,y)~~.
\end{eqnarray}
Defining the central coordinate $X= x+y$ and the relative coordinate $\xi= x-y$, one can rewrite the Boltzmann equation, which is given by:
\begin{equation}\label{kinetic}
  \Big{\{ } \gamma^\mu , i \frac{\partial F(X,q)}{\partial X^\mu} \Big{\}} +  \left[(\slashed{q}-m), F(X,q) \right] = I_{coll}[F] ~~,
\end{equation}
with $p$ the momentum related to the relative coordinates $\xi$. The collision term reads:
\begin{eqnarray}
    &&I_{coll}[F] = - \Sigma^K(X,q) + \Sigma^R(X,q) \ostar F(X,q) - F(X,q) \ostar \Sigma^A(X,q)~~,
\end{eqnarray}
where 
\begin{eqnarray}\nonumber
&& A(X,p) \ostar B(X,p)  =A(X,p) e^{i \left(\frac{\overleftarrow{\partial}}{\partial X_\mu} \frac{\overrightarrow{\partial}}{\partial p^\mu}- \frac{\overleftarrow{\partial}}{\partial p_\mu} \frac{\overrightarrow{\partial}}{\partial X^\mu}\right) } B(X,p) ~~,
\end{eqnarray}

is known as the Moyal product. Rewriting explicitly the self-energy components, assuming that $F(X,p)$ is a scalar function in spinor space, and considering a small dependence on the central coordinates $X^\mu$, one can take only the first term in the expansion of the Moyal product, and one reaches:
\begin{eqnarray}\nonumber
   I_{coll}[F] =  \frac{\gamma m^2}{ m_0^2}  U^\dagger({\bf q}) \Big[ \int dq' g({\bf q-q'})\left( F(X,q')- F(X,q)\right) U({\bf q'})\Delta(q') U^\dagger({\bf q'})\Big] U({\bf q}) ~~,\\
\end{eqnarray}
where  $\Delta(q') = i\mathcal{S}^R(q')-i\mathcal{S}^A(q')$ is the so-called spectral function.  Defining the objects $\mathcal{I}(q) = U({\bf q}) I_{coll}[F] U^\dagger({\bf q})$ and $\mathcal{D}(q) = U({\bf q}) \Delta(q) U^\dagger({\bf q})$  one finds:

\begin{eqnarray}\label{Icoll2}
    &&\mathcal{I}(q) = \tilde{\gamma}  \int dq' g({\bf q-q'})\left( F(X,q')- F(X,q)\right) \mathcal{D}(q')  ~~,
\end{eqnarray}
with $\tilde{\gamma} =\frac{\gamma m^2}{ m_0^2}$.

Consequently,  Eq. \eqref{kinetic} allows us to show that there exists one conserved current $J^\mu$ which can be written as follows:
\begin{eqnarray}
    J^\mu (X) =   \int d q  F(X,q) Tr[\gamma^\mu \Delta(q)]~~,~~ \frac{\partial J^\mu (X) }{\partial X^\mu} = 0~~.
\end{eqnarray}
The energy-momentum tensor is properly given by:
\begin{equation}
  T^{\mu \nu}(X) = \int d q q^\nu F(X,q) Tr[\gamma^\mu \Delta(q)]~~,~~~\frac{\partial T^{\mu \nu}(X)}{\partial X^\mu} = Q^\nu(X)~~,
\end{equation}
and $Q^\nu$ is the heat flow, and is written as:
\begin{equation}
    Q^\nu(X) = \tilde{\gamma}\int dq dq' q^\nu g({\bf q-q'}) \big[F(X,q')- F(X,q) \big] Tr[\Delta(q)\Delta(q')]~~.
\end{equation}

\section{Gap equations}\label{AppD}
Gap equations shown in Eqs. \eqref{GAP1} and \eqref{GAP2} can be written as follows:
\begin{equation}
    \mathcal{Z}({\bf p}) = 1 -\frac{2i}{{\bf p}^2}\frac{\gamma m^2}{m_0^2}  Im\int dp' g({\bf p-p'})Tr[\gamma \cdot {\bf p} \mathcal{S}^R(p') ] ~~,~~ 
\end{equation}
\begin{equation}
B({\bf p}) = m   -2i\frac{\gamma m^2}{m_0^2}  Im \int dp' g({\bf p-p'})Tr[\mathcal{S}^R(p') ]~~,
\end{equation}
with $(\mathcal{S}^R)(p)= \left( \gamma_0 p_0 - \mathcal{Z}({\bf p}) \gamma \cdot {\bf p} - B({\bf p} ) \right)^{-1}$. Thus, one has:
\begin{equation}
    \mathcal{Z}({\bf p}) = 1 -\frac{2i}{{\bf p}^2}\frac{\gamma m^2}{m_0^2}  Im\int dp' g({\bf p-p'})\frac{{\bf p'}^2 \mathcal{Z}({\bf p'} )}{(p'_0)^2 - \mathcal{E}({\bf p'})^2} ~~,~~ 
\end{equation}
\begin{equation}
B({\bf p}) = m   -2i\frac{\gamma m^2}{m_0^2}  Im \int dp' g({\bf p-p'})\frac{B({\bf p'} )}{(p'_0)^2 - \mathcal{E}({\bf p'})^2}~~,
\end{equation}
where $\mathcal{E}({\bf p}) = \sqrt{\mathcal{Z}^2(p) {\bf p}^2 + B({\bf p})^2}$.  Integrating over $p'_0$ one reaches:
\begin{equation}
    \mathcal{Z}({\bf p}) = 1 -\frac{2i}{{\bf p}^2}\frac{\gamma m^2}{m_0^2}  Re \int d{\bf p'} g({\bf p-p'})\frac{{\bf p'}^2 \mathcal{Z}({\bf p'} )}{ \mathcal{E}({\bf p'})} ~~,~~ 
\end{equation}
\begin{equation}
B({\bf p}) = m   -2i\frac{\gamma m^2}{m_0^2}  Re \int d{ \bf p'} g({\bf p-p'})\frac{B({\bf p'} )}{\mathcal{E}({\bf p'})}~~,
\end{equation}
Now, by use of the dimensionless quantities $x =|{\bf p}|/\sqrt{4 \sigma}$ , $y =|{\bf p}|/\sqrt{4 \sigma}$ and integrating over angular variables, one finds:
\begin{equation}
    \mathcal{Z}(x) = 1 -\frac{2i}{x^2}\frac{\gamma m^2}{m_0^2}(4\sigma)  Re \int_0^\infty dy \mathcal{G}(x,y)\frac{y^4 \mathcal{Z}( y)}{ \sqrt{\mathcal{Z}(y)^2 y^2 + B(y)^2/(4\sigma)}} ~~,~~ 
\end{equation}
\begin{equation}
B(x) = m   -2i\frac{\gamma m^2}{m_0^2} (4\sigma)^{3/2} Re \int_0^\infty dy \mathcal{G}(x,y)\frac{y^2 B( y)/(\sqrt{4\sigma})}{ \sqrt{\mathcal{Z}(y)^2 y^2 + B(y)^2/(4 \sigma)}}~~,
\end{equation}
with
\begin{equation}
    \mathcal{G}(x,y) = \frac{1}{\sqrt{2 \pi}}e^{-x^2 - y^2 }\int_0^\pi d \theta  \sin\theta e^{2 x y \cos \theta}  =  \frac{e^{-x^2 - y^2 }}{\sqrt{2 \pi}} \frac{\sinh(2 x y)}{x y}~~.
\end{equation}
\end{widetext}

\end{document}